\documentstyle[aps,multicol,epsf]{revtex}
\def\dir{.}
\begin{document}
\begin{multicols}{2}
\narrowtext
\noindent{\bf Collapse of Randomly Linked Polymers:}
In a recent letter, Bryngelson and Thirumalai (BT)\cite{BT} consider
an ideal (i.e. non--self--interacting) polymer in which $M$ randomly
chosen pairs of monomers are constrained to be in close proximity.
The unconstrained chain of $N$ monomers is expanded, with a 
typical radius scaling as $R\propto \sqrt{N}$. By comparing variational 
estimates to the free energies of expanded and collapsed states, 
BT argue that increasing the number of (uncorrelated) links causes 
the polymer to collapse into a localized state in which $R$ is 
independent of $N$. In particular, they conclude that for a generic 
set of constraints, where the typical distance $\ell$ 
(measured along the backbone) between linked monomers is 
of the order $N$, a negligible density of constraints 
($\sim 1/\ln N$) can cause such a collapse. 

Here, we demonstrate that the polymer remains expanded
unless $M\sim N$.
Rather than concentrating on estimates of the free energy,
we measure directly the squared end to end distance $r^2$, 
of the polymer. We derive an exact lower bound $r^2>N/M$,
which proves that, contrary to the conclusion of BT, uncorrelated 
links do not cause the polymer to collapse. Numerical 
simulations are also performed by exploiting an 
analogy to random resistor networks\cite{RRN}: 
The squared end to end distance corresponds to the resistance 
of a chain of uniform resistivity in which randomly chosen pairs
of points are connected by shorts of zero resistance. Simple
elimination of series and parallel resistors reduces the problem
to a network of at most $M$ nodes. Extensive simulations
show that $r^2\approx 1.5N/M$.

The lower bound for $r^2$ is obtained by noting that the $M$ links
break the polymer into $2M+1$ segments. $r^2$ is 
certainly larger than the sum of the end to end distances of the 
two extremal segments. (The resistance of the network is larger
than its two end pieces.) In the limit of large $M$, the length of
each segment is independently taken from an exponential 
distribution with a mean size of $N/(2M+1)$. We choose
parameters such that a segment of length $s$ has squared end
to end distance of $r(s)^2=s$, i.e. the corresponding chain 
has unit resistivity. The contributions of 
the two end segments thus add up to $2N/(2M+1)\approx N/M$. 
Therefore, for large $M$ we have $r^2>N/M$. This bound ensures
the absence of a localized state for finite density $M/N$. 

\begin{figure}
\epsfysize=3.0truein
\vbox{%\vskip 0.15truein 
\hskip 0.4truein
\epsffile{\dir/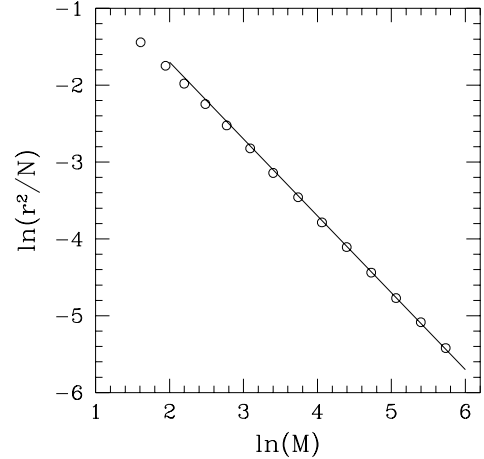}
%\vskip 0.15truein
\caption{Logarithmic plot of the scaled squared end to end distance
as a function of number of links $M$. Each point is an average 
over 1600 randomly linked chains. The solid line represents
$r^2/N=1.5/M$.}
\label{FigOne}
}
\end{figure}

A simple scaling argument suggests that in the continuum limit of
$N\gg1$, $r^2({M,\lambda N})=\lambda r^2({M,N})$, and hence
$r^2=f(M)N$. We confirmed numerically this scaling  
for several values of $N$. Fig.~\ref{FigOne} depicts (on a logarithmic 
scale) $r^2/N$, as a function of $M$, 
for a single value of $N=2560$. Every point on this figure represents 
an average  over 1600 configurations of random links. The numerical 
results gradually converge to a slope of -1, as depicted by the solid line.
A least squares fit to {\em all} points of the figure produces a 
slope 0.97, and the curve cannot be fitted as $\ln M/M$. In fact,
we conclude that $r^2\approx 1.5N/M$; with a prefactor that is 
surprisingly close to the value of 1 which appears in our simple 
lower bound.

Details of the numerical algorithm, as well as a discussion of implications
for  self--avoiding polymers appear in a companion paper\cite{KK}. 
After completion of this work, we became aware of a preperint by 
Solf and Vilgis\cite{SV}. Although they consider more
general polymer networks, their results also
agree with the above findings.

We acknowledge support by the US--Israel BSF grant 
No.~92--00026 and by the NSF grant No.~DMR--94--00334.

\date{\today}
\pacs{35.20.Bm, 36.20.--r, 64.60.--i, 87.15.By.}
\noindent Yacov Kantor$^1$ and Mehran Kardar$^2$

$^1$School of Physics and Astronomy
\par Tel Aviv University
\par Tel Aviv 69 978, Israel
\par$^2$Department of Physics
\par Massachusetts Institute of Technology
\par Cambridge, Massachusetts 02139

\end{multicols}
\end{document}